\documentclass[journal,twoside]{IEEEtran}

\usepackage[normalem]{ulem}
\usepackage[table,xcdraw]{xcolor}
\ifCLASSINFOpdf
  \usepackage[pdftex]{graphicx}
  \graphicspath{{./figs/}}
  \DeclareGraphicsExtensions{.pdf,.png}
\else
  \usepackage[dvips]{graphicx}
  \graphicspath{{./figs/}}
  \DeclareGraphicsExtensions{.eps,.mps}
\fi
\usepackage[cmex10]{amsmath}

\ifCLASSOPTIONcompsoc
 \usepackage[caption=false,font=normalsize,labelfont=sf,textfont=sf]{subfig}
\else
 \usepackage[caption=false,font=footnotesize]{subfig}
\fi
\usepackage{dblfloatfix}

\usepackage{url}
\usepackage{textcomp}
\usepackage{enumitem}
\usepackage{xcolor}
\usepackage{comment}
\usepackage{lineno}
\usepackage{eurosym}
\usepackage{tabularx}
\usepackage{booktabs}
\usepackage{multirow}
\usepackage{makecell}
\usepackage{pbox}
\usepackage{psfrag}

\newcommand{\ti}[1]{\textcolor{black}{\textit{#1}}}

\usepackage{float} 
\usepackage[abs]{overpic} 

\usepackage{psfrag} 

\usepackage{multicol}

\newcommand{\fig}[1]{Fig.\,\ref{#1}}
\newcommand{\equ}[1]{Eq.\,(\ref{#1})}
\newcommand{\tab}[1]{Table\,\ref{#1}}

\newcommand{\ph}[1]{\phantom{#1}}

\usepackage{comment}
\excludecomment{hide} 

\begin{document}


\title{Exploiting the Solar Energy Surplus\\ for Edge Computing}

\author{
    Borja~Martinez,~\IEEEmembership{Senior~Member,~IEEE,}
	Xavier~Vilajosana,~\IEEEmembership{Senior~Member,~IEEE}%
    \thanks{B.\,Martinez is with IN3 at Universitat Oberta de Catalunya.}
	\thanks{X.\,Vilajosana is with IN3 at Universitat Oberta de Catalunya and Worldsensing S.L.}%
}



\begin{hide}
\ifCLASSOPTIONpeerreview
  \markboth{IEEE Transactions on Sustainable Energy}%
  {Exploiting the Solar Energy Surplus for Edge Computing}
\else
  \markboth{IEEE Transactions on Sustainable Energy}%
  {Exploiting the Solar Energy Surplus for Edge Computing}
\fi
\end{hide}

\maketitle

\begin{abstract}


In the context of the global energy ecosystem transformation, we introduce a new approach to reduce the carbon emissions of the cloud-computing sector and, at the same time, foster the deployment of small-scale private photovoltaic plants. We consider the opportunity cost of moving some cloud services to private, distributed, solar-powered computing facilities. To this end, we compare the potential revenue of leasing computing resources to a cloud pool  with the revenue obtained by selling the surplus energy to the grid. We first estimate the consumption of virtualized cloud computing instances, establishing a metric of computational efficiency per nominal photovoltaic power installed. Based on this metric and characterizing the site's annual solar production, we estimate the total return and payback. The results show that the model is economically viable and technically feasible. We finally depict the still many questions open, such as security, and the fundamental barriers to address, mainly related with a cloud model ruled by a few big players.

\end{abstract}


\IEEEpeerreviewmaketitle

\section{Introduction}
\label{sec:introduction}

\IEEEPARstart{E}{nergy}
is still today primarily generated from non-renewable sources, although the trend has been changing in the last few years. Green energies are progressively being adopted, and renewable power generation is now growing faster than the overall power demand~\cite{IRENA-2020}. Sustaining this growth, energy generation is being spread not only in utility-scale mega plants, but also promoting small-scale projects through residential, schools, factory roofs and other non-commercial deployments.

These geographically distributed, small-scale plants are lately being encouraged by the drop in the costs of solar technologies and the sustained increase in electricity utility rates. 
The confluence of these two trends (whose intersection depends on the energy prices, but also on the incentive policies and taxes of each country) ensures that self-generation can be a cost-effective option at some time.
%
In addition, policy makers are introducing new regulations to favor producers that generate electricity through their own private resources, 
so that they are paid back by the utility companies for their energy surplus when injected into the grid. 
%
However, it is still not clear which pricing model utility companies will adopt. 
If companies are only required to pay back the energy at a market price, that is, set the price according to the offer and demand at the production time, 
the price will eventually drop to zero during peak production hours. 

Creating a fair market means being able to diversify. 
That is, to give PV producers an alternative to sell the energy at peaks production, which may force them to accept very low prices, even zero. 
The apparently obvious solution would be to store the energy locally until a favorable time for sale. 
However, the price of large-capacity batteries is still very high (the payback is not guaranteed), 
and it is not clear that prices will go down soon, since other sectors such as the electric vehicle are creating a huge demand.

In this article we explore the viability of an alternative use of the energy by enabling local computation on-demand. 
This is materialized by exploiting the energy to power computing hardware attached to the local grid, leasing resource slots for third-party computation.
The local cluster may be seen as an edge extension of a cloud infrastructure operating on the basis of self-generated energy. 
This possibility opens up a whole new approach, a way to give more value to energy while responding to the world's computing needs, by moving this computing to where the energy is produced.

The article is organized as follows, Section \ref{sec:overview} introduces the model and architecture of the proposed solution and presents our prototype implementation used to validate the technical feasibility. 
Section \ref{sec:computing} introduces the computing model used to estimate the potential revenues derived from computation. 
Section \ref{sec:solar} analyzes our capability to forecast the production of energy to guarantee the availability for the services offered
and Section \ref{sec:viability} puts it all together combining revenues obtained from the computing model and the capability to run it with the energy surplus. 
Section \ref{sec:sota} then presents the state of the art, identifying ongoing research activities and initiatives aligned to the ideas presented in this article. 
Finally, Section \ref{sec:conclusion} adds some closing remarks and discusses open-ended questions of the model.

\section{Model and Architecture Overview}
\label{sec:overview}

\begin{figure}[ht!]
  \centering
  \includegraphics[width=1.0\columnwidth]{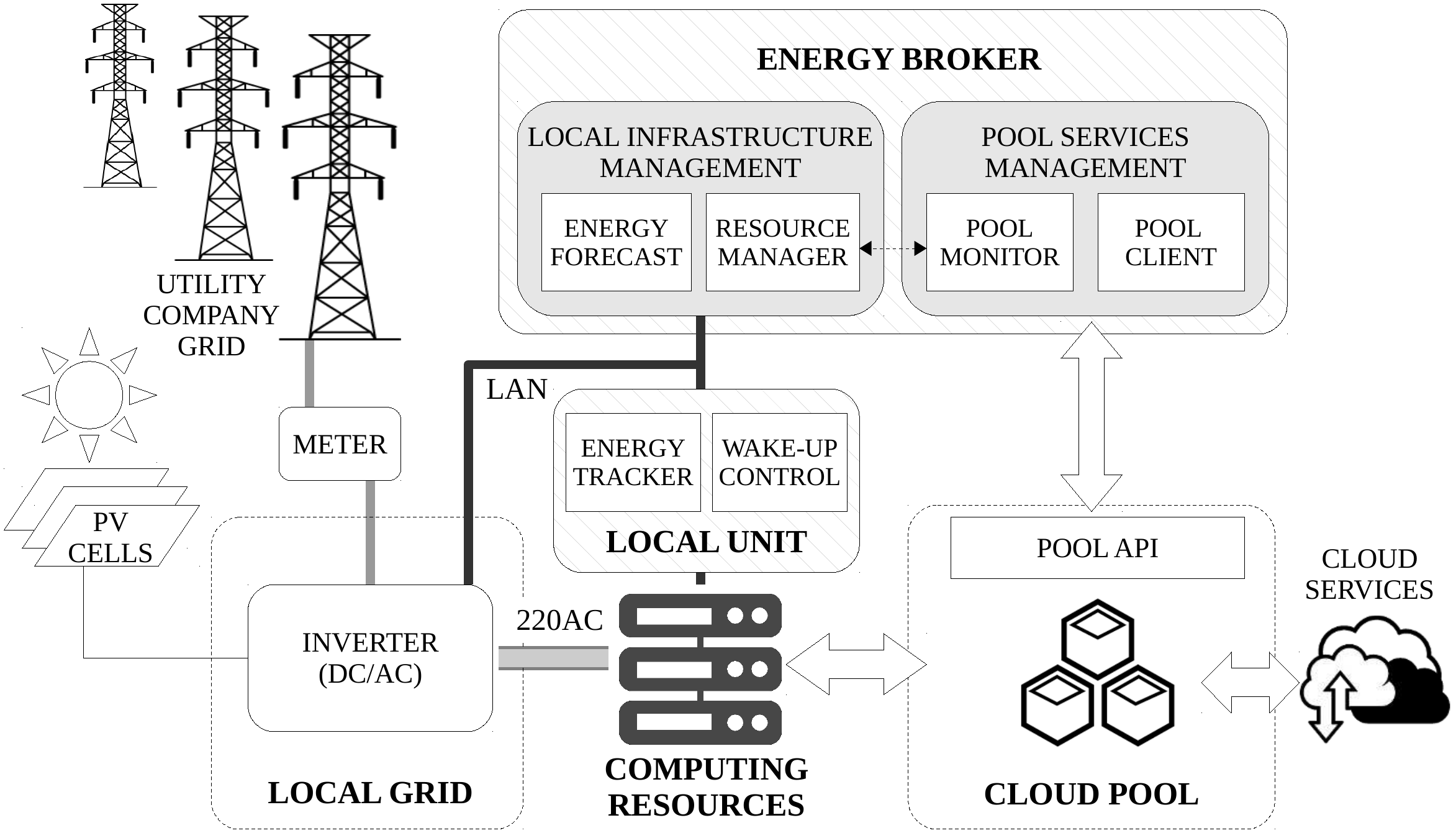}
  \caption{Architecture of the green computing model in which a local PV system is connected to computing equipment and operated according to the energy supply and computing demand by a Resource Broker. The Resource Broker has the capability to forecast energy production as well as monitor demand for computation from different cloud computing sources.}
  \label{fig:f1}
\end{figure}

In this section we present the architecture of the envisaged solution and provide some details about our prototype implementation.
\fig{fig:f1} presents a schematic diagram of the proposal, in which PV cells are installed in some local facility (e.g. a domestic installation). 
The energy is forward to a DC/AC converter (commonly referred as inverter) which is placed in the \ti{local grid}.
In its turn, the inverter can inject the energy to the utility company grid if not consumed locally. 
The system also comprises a set of \ti{computing resources} deployed in the local premises (e.g. a small rack with a few computing blades). 
An additional hardware unit (referred as \ti{local unit} -LU-) is attached to the local network, being able to monitor the activity in the inverter (\ti{energy tracker}) as well as having control of the computing activity (\ti{wake-up control}).

A software system running on the cloud (referred to as \ti{energy broker} -EB-) is able to trigger the activation of the computing equipment (\ti{resources manager}) when certain conditions are met,
which is carried out through the LU. 
The EB is able to forecast the local energy production (\ti{energy forecast}), monitoring the energy price and also evaluating the price that is being paid in different computing services providers. 
With this information the broker may decide at a certain point to turn on the computing facilities and join them to a computing pool (\ti{pool client}), announcing the computation availability. 
Obviously, that will only be done if the estimated revenue obtained by computing during a predefined period of time is higher than that of selling the excess energy to the utility company. 
While computing resources are leased to a pool, accounting of its usage is performed by the \ti{pool monitor}. 

The described architecture have been implemented so as to validate the technical feasibility of such model. 
The software implementation has been conducted in Python and followed a distributed micro-services architecture, interconnected through a MQTT broker. 
The software components are distributed in two blocks, the EB software, running on the cloud and the LU software operating in the local premises network.

The LU has been implemented using a Raspberry PI 3b+. 
In it, a software agent is in charge of interacting with the online EB and has the capability to turn on/off and communicate with the computing resources (in our case implemented with a rack of Raspberry PIs).
 
The LU uses the \textit{wake-on-lan} functionality to turn on some of the computing nodes, which upon start run a software agent that connects to a local MQTT broker hosted at the LU. 
The LU acts as a proxy with the online EB who manages the computing infrastructure according to the forecasts. 
The LU is also connected to the API of the inverter, which monitors in real time photovoltaic generation and local energy consumption.
This information is summarized and reported to the online energy broker in a regular basis. 
 
The LU also reports information about the status of the computing facilities, for example indicating whether they are active or stopped and the status of the running software if any. 
This information can be used to assess the utilization of the on-premises computing facilities, serving also as a tool to match the usage reported by the pool monitor.

The EB software is an online, cloud-hosted micro-service, that has an holistic view of different LUs. 
It stores information about the generated energy in each of the connected infrastructures 
as well as the status of the computing facilities attached. 
The energy broker is also connected to different weather services (Accuweather \cite{accu-api}). It is, as well, constantly forecasting for each of the LUs connected to it the energy availability for the next hour. In addition it is querying different pools for computation demand and pricing. 
In our prototype, we used \ti{SETI@home} \cite{Korpela2001} to emulate computing demand. 
This is, upon forecasting some energy availability, the energy broker notifies the LUs that may have energy available. 
The LUs wake up the required computing resources to service that demand by executing the BOINC agent \cite{Boinc2004}, attaching it to the \ti{SETI@home} project. 
This particular pool does not provide any revenue, but the same principle can be applied to launching a Kubernetes instance through KubeEdge, launching the AWS Greengrass or the Azure IoT agents to receive computation on demand. 

\section{Computing Model}
\label{sec:computing}

In our aim to understand the potential revenues from our approach, we need to estimate the power consumption of the computing resources and map the energy to its computing value. 
To this end, we defined a normalization parameter $\eta_C$ to account for the number of computing units that can operate per kW. 
With this definition, $R_C$ in \equ{eq:eq1} represents the reward per kWh of energy that can be obtained by leasing computing resources, considering an hourly allocation price V$_I$ for each computing instance:

\begin{equation}
\label{eq:eq1}
R_C [\text{\texteuro}/kWh] = \eta_C [1/kW] \cdot V_I [\text{\texteuro}/h] \cdot  \alpha
\end{equation}

In this equation $\alpha$ is the allocation factor that accounts for the resources that are actually leased from those that are offered. That may include cases in which there is not enough computation demand to load all offered instances or 
instances that are leased only a fraction of the available time.
This value may be included as part of the negotiation process with the pool. 

To estimate the value of $\eta_C$ we followed an empirical approach by measuring the power consumption of real, state-of-the-art computing instances. 
To that goal, we followed the methodology taken by Khan \textit{et al.}\cite{khan19rapl} in measuring the power of cloud computing instances. 
In their article, the RAPL Intel Software power meter \cite{rapl} and a load generation tool \cite{Hirki16desc} are used for that aim. 

Following the same methodology, the power consumption of different Amazon Web Service (AWS) instances has been measured using, in our case, the \textit{strees-ng} \cite{stress-ng} processor-load generation tool and the RAPL toolset. 
As reported by the same authors, there exists uncertainty in the measurements because cloud instances are shared: 
the different cores of an actual HW processor can be assigned to multiple virtual instances, and therefore measuring the energy consumption in such instances leads to imprecise results. To understand this uncertainty the measurements have been conducted in different AWS instances, including those in which the entire processor is assigned.

A second source of uncertainty is that, in a virtualized environment, the actual instance assigned can vary from launch to launch without control from the customer side. 
In fact, the processors' features can actually be very different (vendor, model, generation, etc.) provided that they meet some high-level constraints, 
as those reported in \tab{tab:aws-instances}. 

In our experiments, we selected the T2 series from AWS, a family of low-cost, general purpose instances. Common processors assigned to that family (referred as Amazon Lightsail) include the 8 core Intel(R) Xeon(R) CPU 5-2676 v3 and the E5-2686 v4 family. 
They both show similar computing performance but different energy consumption characteristics, being the latter significantly more power hungry.

\begin{figure}[ht!]
  \centering
  \includegraphics[width=1.0\columnwidth]{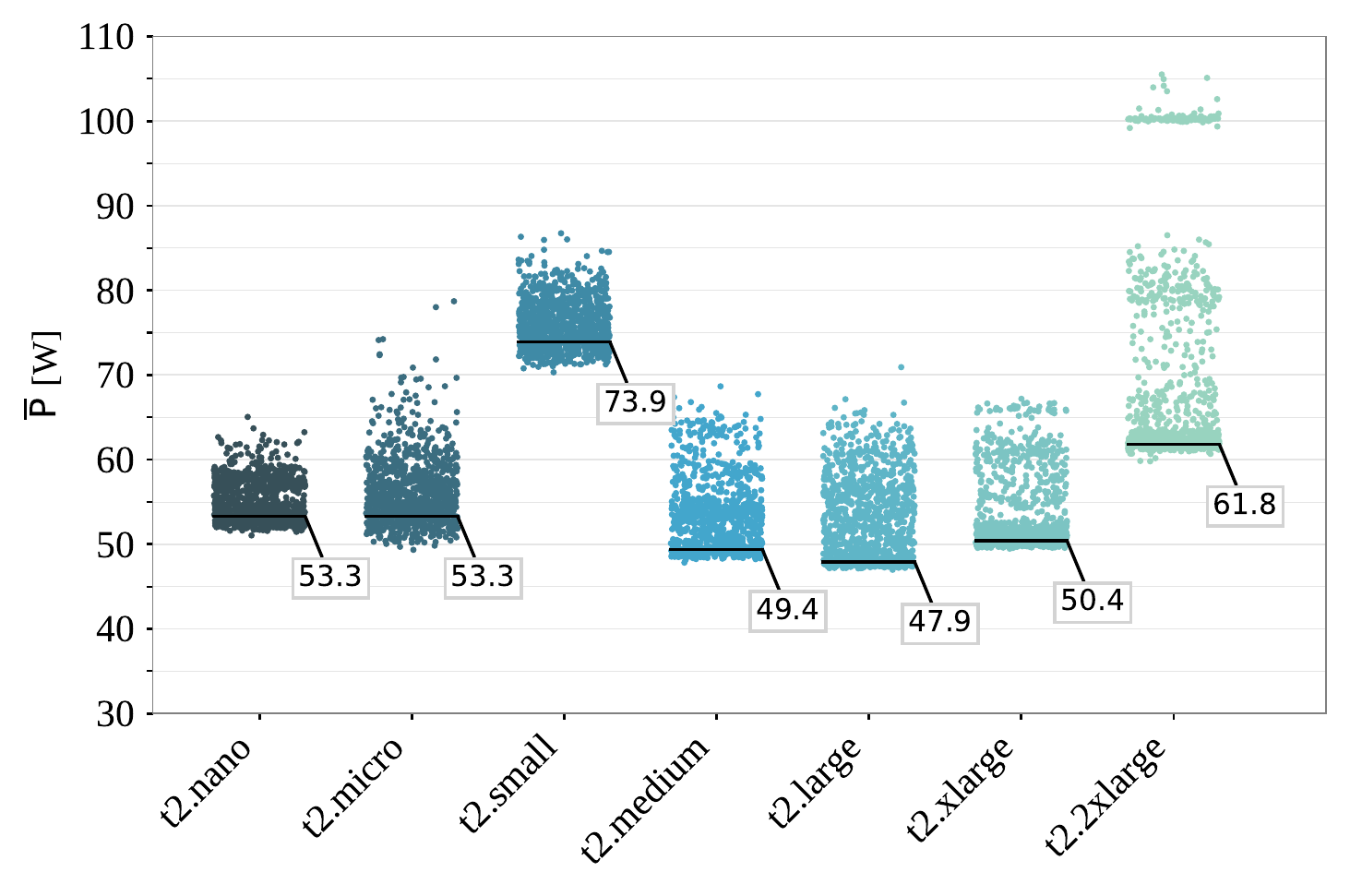}
  \caption{Distribution of the measured power consumption in AWS T2 instances with 100\% load of the instance cores. The statistical mode for each of the measured instances is annotated in the figure.}
  \label{fig:f2}
\end{figure}

\fig{fig:f2} presents the power consumption measured on each of the selected AWS T2 instances when the stress-ng is forcing each of the instance cores to 100\% of the capacity. 
The figure collects 1000 measurements for each of the instances in order to ensure statistical significance. 
As can be observed, the power consumption of an instance is not tightly correlated to the number of cores and available RAM memory (summarized in \tab{tab:aws-instances}). 
Instead consumption is highly dependent on the underlying processor characteristics. 
For example, we see very similar power consumption for a T2.micro and T2.xlarge, even when the former is using a single core while the latter is using 4 of them. 
With these results we corroborate the uncertainty in measuring AWS instances as described in \cite{khan19rapl}. 
Nevertheless, we take these results as tentative bounds for the power consumption of an instance, 
giving us an indication of what is the power consumption range we may expect for a typical computing platform. 
Specifically, derived from the experimental measurements, we take a range of 50W to 100W per instance along the remainder of the article.

\begin{table}[ht]
\centering
\caption{AWS T2 Instances characteristics and leasing price (2020).}
\label{tab:aws-instances}
\setlength{\tabcolsep}{10pt}
    {
    \begin{tabular}{p{1.5cm} c c c c}
    \toprule
    \multicolumn{1}{l}{\textbf{Name} } & 
    \multicolumn{1}{c}{\textbf{vCPU} } & 
    \multicolumn{1}{c}{\textbf{RAM} } & 
    \multicolumn{1}{c}{ $\eta_C$ } & 
    \multicolumn{1}{c}{ $V_I$ } \\
    \midrule
    t2.nano    & 1 &      0.5  & 18.8 & 0.0059\\
    t2.micro   & 1 & \ph{0.}1  & 18.8 & 0.0118\\
    t2.small   & 1 & \ph{0.}2  & 13.5 & 0.0236\\
    t2.medium  & 2 & \ph{0.}4  & 20.2 & 0.0472\\
    t2.large   & 2 & \ph{0.}8  & 20.9 & 0.0944\\
    t2.xlarge  & 4 & \ph{.}16  & 19.8 & 0.1888\\
    t2.2xlarge & 8 & \ph{.}32  & 16.2 & 0.3776\\
    \midrule
    \multicolumn{5}{l}{$\eta_C$ based on the most common value (mode) of the measurements.}\\ 
    \multicolumn{5}{l}{$V_I$ Price for instance from AWS (2020) as for reference}\\
    \bottomrule
    \end{tabular}%
    } 
\end{table}


\section{Solar Model}
\label{sec:solar}
The next step to validate the viability of the proposed model involves the understanding of the solar energy generation patterns. 
For this particular application, a PV model needs to consider two aspects:
\begin{itemize}
\item Estimate the aggregated energy generation throughout the year that will determine the annual payback, and thus the system's viability.
\item From an operational point of view, it is necessary to know in advance the solar availability for the next short-term time slot. 
This is relevant to determine whether the computer facilities should be launched or announced as available to the pool.
\end{itemize}

\subsection{Aggregated Energy Model}
In order to calculate the accumulated energy produced it is convenient to use the concept of equivalent peak-sun hours (PSH). PSH is a method to simplify the handling of the cyclic variations of solar irradiance (day/night, seasons,...) and other external factors (such as local weather conditions, temperature, etc).  
Technically, a peak sun-hour is an hour during which the intensity of sunlight is maximal, which is considered around 1 $kW/m^2$ 
(specifically, STC conditions are defined as 1 $kW/m^2$, at 25$\,C^\circ$ and Air-Mass 1.5).  
The insulation, defined as the solar energy that is incident on a specified area/object over a period of time (e.g. daily, monthly, annually), 
can be expressed as the energy that would be captured at full irradiance (STC) during an equivalent time period of PSH hours. 
The actual PSH of a given location is determined by the solar cycles at the geographical position, but also by meteorological phenomena preventing rays from reaching the earth's surface, such as clouds and rain. 
The PSH of a site is, therefore, a measure of the accumulated plane of array irradiance (POA), that is, the actual incident irradiance reaching the cell array. 
PSH is a very convenient unit because the solar panels specifications are given at STC conditions, that is, full irradiance. 
Then, from the definition of PSH, the total energy produced by a PV system with nominal power $P_{MPP}$ can be expressed as: 

\begin{equation}
\label{eq:eq2}
E_T\,[kWh] = PSH \cdot P_{MPP} \cdot \eta_{SYS}
\end{equation}

In \equ{eq:eq2}, the efficiency factor $\eta_{SYS}$ accounts for the PV system losses (such as ohmic losses in wires, DC/AC inverters, etc) and capture losses (mainly the impact of temperature in the silicon efficiency and reflectance effects). 
Typical system losses are around 15\% \cite{Marion-2005}. 
Regarding capture loss, reflectivity entails an additional 2-4\%, 
but the cell-temperature loss can be very different depending on the placement, sometimes very high \cite{Huld-2015}.
It should therefore be estimated specifically for the particular site. 
High pollution in some specific areas can also impact negatively. 

Starting from 2001, the Joint Research Centre of the European Commission developed the well established open source framework named 
Photo-Voltaic Geographical Information System (PVGIS) \cite{Suri-2005} \cite{Huld-2005}. 
In essence, PVGIS is a tool-suite to estimate the insulation in large geographical regions worldwide based on satellite observations \cite{Mueller-2009}, 
and it has become a widespread research tool for the performance assessment of PV plants. 

\begin{figure*}[ht!]
  \centering
  \includegraphics[width=1.0\textwidth]{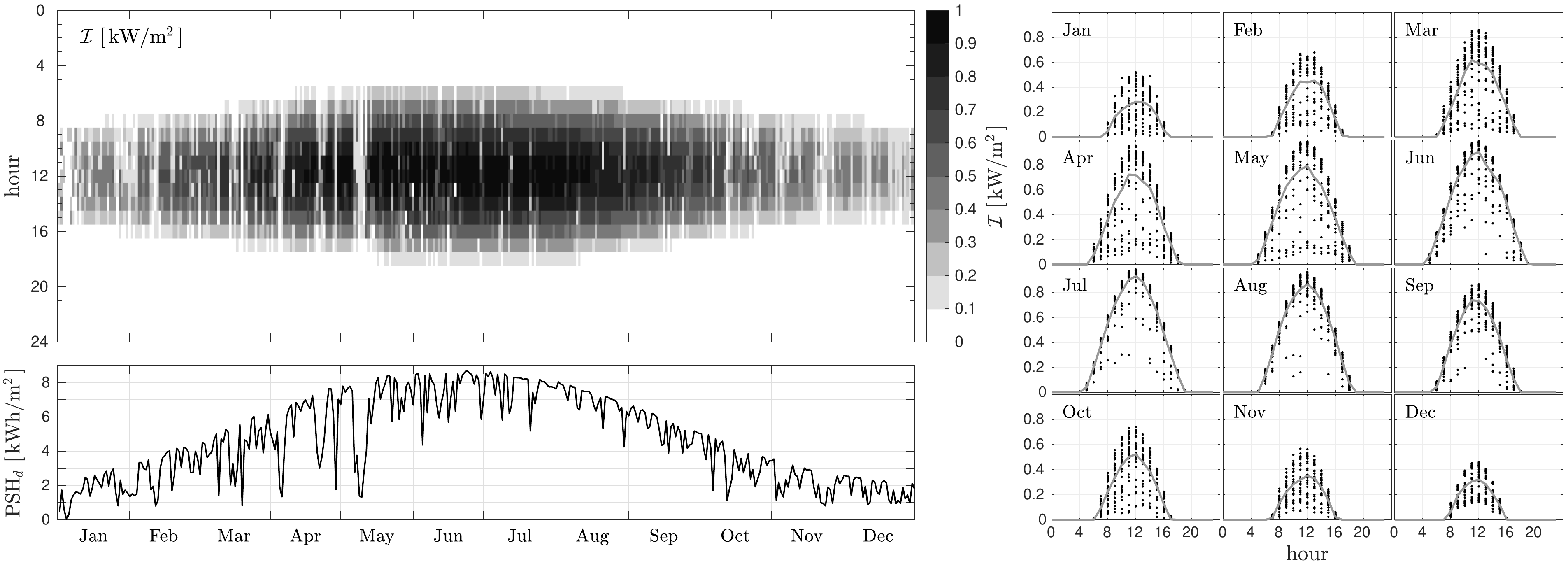}
  \caption{Annual irradiation evolution. Top-left, evolution of irradiation in relation to the hour of the day along the year. Bottom-left, evolution of the equivalent Peak Sun Hours (PSH) per day along a year. Right, irradiation variability per hour, grouped by month.}
  \label{fig:f3}
\end{figure*}

PVGIS outputs the in-plane solar irradiation and PV production with different time aggregations, 
detailing the losses in the PV output caused by various effects discussed above. 
For instance, \fig{fig:f3} (left) presents the hourly-based irradiation at 0$^\circ$ (horizontal plane) for the year 2016, obtained from PVGIS. 
As can be observed, the solar daily and seasonal patterns are readily identifiable, despite the abundant short-term periods with significant deviations from the irradiation pattern,
which are attributable to local transitory weather conditions.

\begin{table}[ht]
\centering
\caption{Yearly PSH and PV generation}
\label{tab:t2}
\setlength{\tabcolsep}{10pt}
    {
    \begin{tabular}{p{3cm} c c}
    \toprule
    \multicolumn{1}{l}{\phantom{0}} & 
    \multicolumn{1}{c}{\textbf{Tilt=0$^\circ$} } & 
    \multicolumn{1}{c}{\textbf{Tilt=30$^\circ$} } \\
    \midrule
    PSH ($\mu \pm \sigma$)\,$\dagger$  & 1670.7 $\pm$ 38.3 & 1968.8 $\pm$ 52.2  \\
    PV Generation\,$\ddagger$                & 1293.0 $\pm$ 27.4 & 1543.4 $\pm$ 39.5  \\
    Total Loss $\eta_{SYS}$\,$\mathsection$    & 22.61\%            & 21.60\%  \\
    \midrule
    \multicolumn{3}{l}{$\dagger$ Peak-sun-hours [kWh/m$^2$] per year. Location +41.53$^{\circ}$N, 2.23$^{\circ}$E}\\ 
    \multicolumn{3}{l}{$\ddagger$ Total production per year [kWh] for a nominal power P$_{\text{MPP}}$=1kW.}\\
    \multicolumn{3}{l}{$\mathsection$ PVGIS system loss input = 15\%}\\
    \bottomrule
    \end{tabular}%
    } 
\end{table}

\tab{tab:t2} presents the annual irradiation (cumulative irradiance) and its variability (mean $\mu$ and standard deviation $\sigma$) in a period of 12 years (2005-2016), also obtained from PVGIS. 
As can be observed, in the aggregation the variability is low ($\sigma$ relatively small as compared to the average $\mu$), despite the seemingly randomness of holes appearing in \fig{fig:f3}.

The total PV production (kWh/year), once priced in \texteuro{€}/kWh, is the key parameter to compute the payback of the equipment and possible revenues, as we will show in Section \ref{sec:viability}. 
Due to this characteristic small variability, the payback can be considered as deterministic in terms of production, that is, it can be calculated with high confidence for our purposes. 

\subsection{1-Hour Ahead Model}
\begin{figure}[ht!]
  \centering
  \includegraphics[width=1.0\columnwidth]{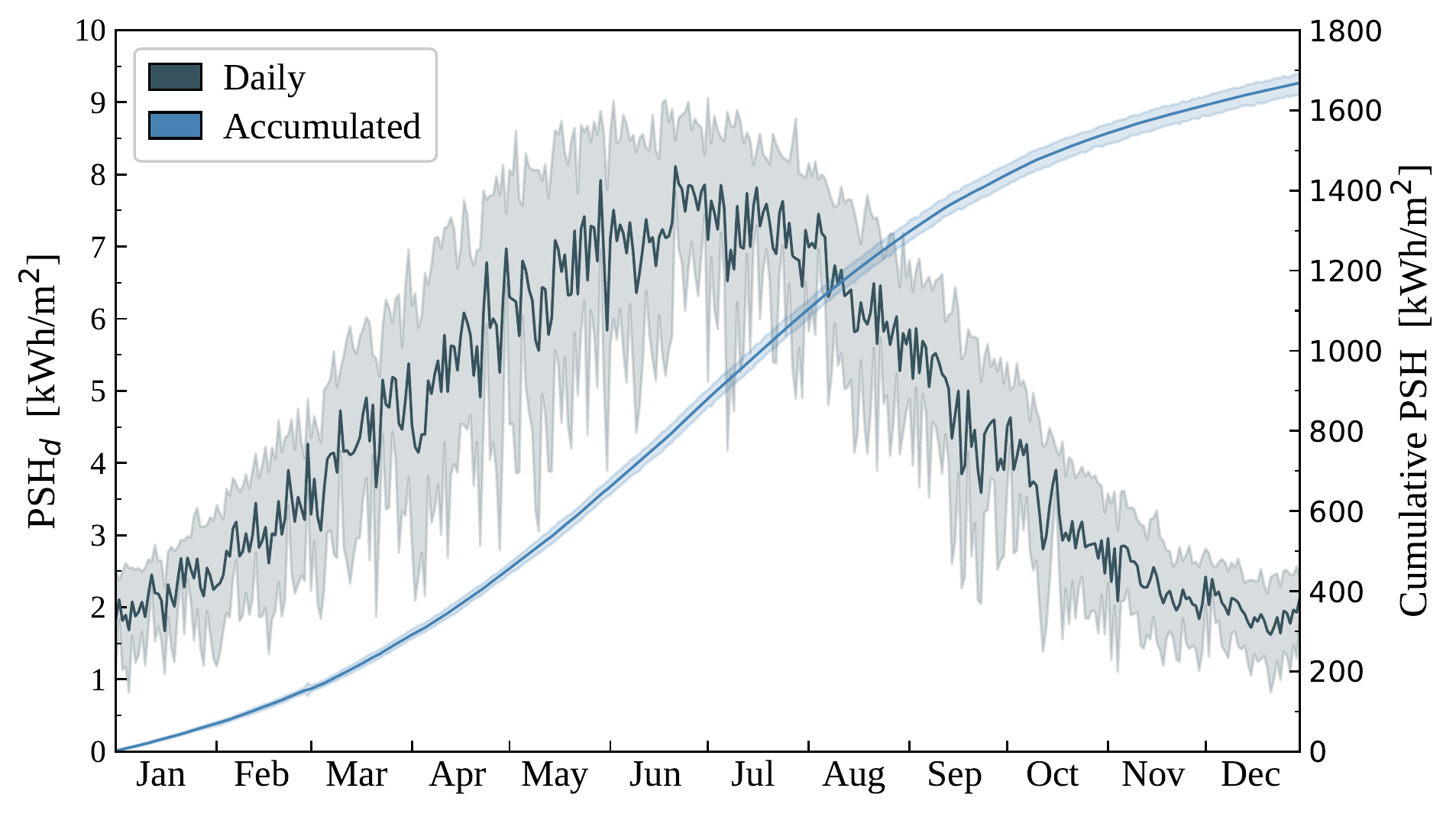}
  \caption{Irradiation variability: Daily PSH irradiation (left axis) and accumulated (right axis). Solid line indicates the mean value for the period (2005-2016) and the shaded area represents the 95\% confidence interval.}  
  \label{fig:f4}
\end{figure}

To analyze with more detail the variability of the solar irradiation we collected data from PVGIS on a daily basis for the same 12 year period.  
\fig{fig:f4} details the daily irradiation (left) and the accumulated evolution (right), expressed in PSH. 
As can be observed, the daily-based variability over the years is significant (due to changing weather conditions) but the accumulated values tend to rapidly cancel these fluctuations, resulting in the small variability reported in \tab{tab:t2}. 

From another angle, \fig{fig:f4} (right) presents the hourly variability of the solar irradiation by month. 
Each point in the tiles corresponds to a particular observation (irradiation) at that specific hour in different years. 
The solid lines represent the mean value in the record at that specific hour. 
As can be observed, the variability in an hourly basis is large in most of the seasons, even showing gaps with almost null irradiation. 

Certainly, PV production will depend on the amount of irradiation received, and this determines the number of instances that can be powered by the energy generated. 
Thus, in our model, an important factor to consider is the radiation expected in the immediate, foreseeable future; looking at these short-term spots on which PV drops significantly.
It is therefore desirable to know in advance -with some guarantees- whether the next computing slot (e.g hour) the PV sub-system will provide an energy surplus. 
To that goal, we propose a "1-hour ahead" forecast model for PV production, based on current weather observations and short term weather forecast. 


Note also that the model should be independent of the particular parameters of an specific PV installation, 
relying only on variables that impact production in similar way for all deployments and are accessible at each deployment, 
discarding information that may be only available at some plants (e.g. some specific output of a particular vendor inverter).

In that direction we evaluated a regression model that exploits different sources of information. Those include:

\begin{enumerate}
\item \textbf{Solar irradiation:} Maximum irradiation in a clean atmosphere given the position of the sun, which is determined by geographic coordinates (latitude/longitude), day of the year, hour and tilt of the panels. In our case we used the well-known library \cite{pvlib-2018}.
\item \textbf{Historical data:} aggregated records, which take into account local climatic phenomena (clouds, rain, etc.), from PVGIS. 
\item \textbf{Meteorological information}: current conditions (which are highly correlated to current production) and short term forecast (temperature, humidity, cloud cover, precipitation, etc.). 
Specifically, we used the Accuweather API \cite{accu-api}, as it is a well known reference for local weather forecasts. Other sources may have worked for our purpose. 
\end{enumerate}

\tab{tab:t3} summarizes the particular explanatory variables that have been considered to derive the predictive model.

\begin{table}[ht]
\centering
\noindent
\caption{Yearly PSH and PV generation}
\label{tab:t3}
\setlength{\tabcolsep}{3pt}
    {
    \noindent
    \begin{tabularx}{\columnwidth}{p{2.2cm} X l}
    \toprule
    \textbf{Source} & 
    \textbf{Description} &
    \textbf{Variables} \\
    
    \midrule
    \parbox[t]{\linewidth}{PVLIB Model} &
    {Power predicted based on irradiation given by solar position at specific location and daytime.} &
    \setlength{\tabcolsep}{0pt}
    \begin{tabular}[t]{l}
            Latitude\\
            Longitude\\ 
            Date-Time\\
    \end{tabular} \\
    
    \midrule
    \parbox[t]{\linewidth}{AccuWeather API:\\Observation} &
    {Last weather observation.
     "Weather Text" is a tag synthesizing the current conditions (Sunny, Cloudy, etc.).} &
    \setlength{\tabcolsep}{0pt}
    \begin{tabular}[t]{l}
            Cloud Cover\\
            Temperature\\ 
            Wind Speed\\
            UV-index \\
            Weather Text\\
    \end{tabular} \\
    
    \midrule
    \parbox[t]{\linewidth}{AccuWeather API:\\Forecast} &
    {Forecasted weather conditions for the following hour.} & 
    \setlength{\tabcolsep}{0pt}
    \begin{tabular}[t]{l}
            Cloud Cover\\
            Temperature\\ 
            Wind Speed\\
            Precipitation\\
            Humidity\\
    \end{tabular} \\
    
    \bottomrule
    \end{tabularx}%
    } 
\end{table}

For the validation of the model, 
we used a dataset of four months of production with hourly traces from a mid-size PV plant, located in a factory flat roof at latitude 41.4$^\circ$ (Barcelona area). 
The system has a nominal power of 380\,kWh, with panels installed at tilt 0$^\circ$ and azimuth 180$^\circ$ (south faced). 

In our experiments, we evaluated different simple regression approaches to find a baseline in the achievable accuracy of the forecast: 
Generalized Linear Models (GLM), 
Ridge, 
Bayesian Ridge,
Least Absolute Shrinkage and Selection Operator (LASSO) 
and Support Vector Machine - Regression (SVR).
 
In our case the best accuracy has been obtained using the SVR method, but we would like to remark that further optimizations can be undoubtedly achieved with more advanced methods.

\begin{figure*}[ht!]
  \centering
  \includegraphics[width=1.0\textwidth]{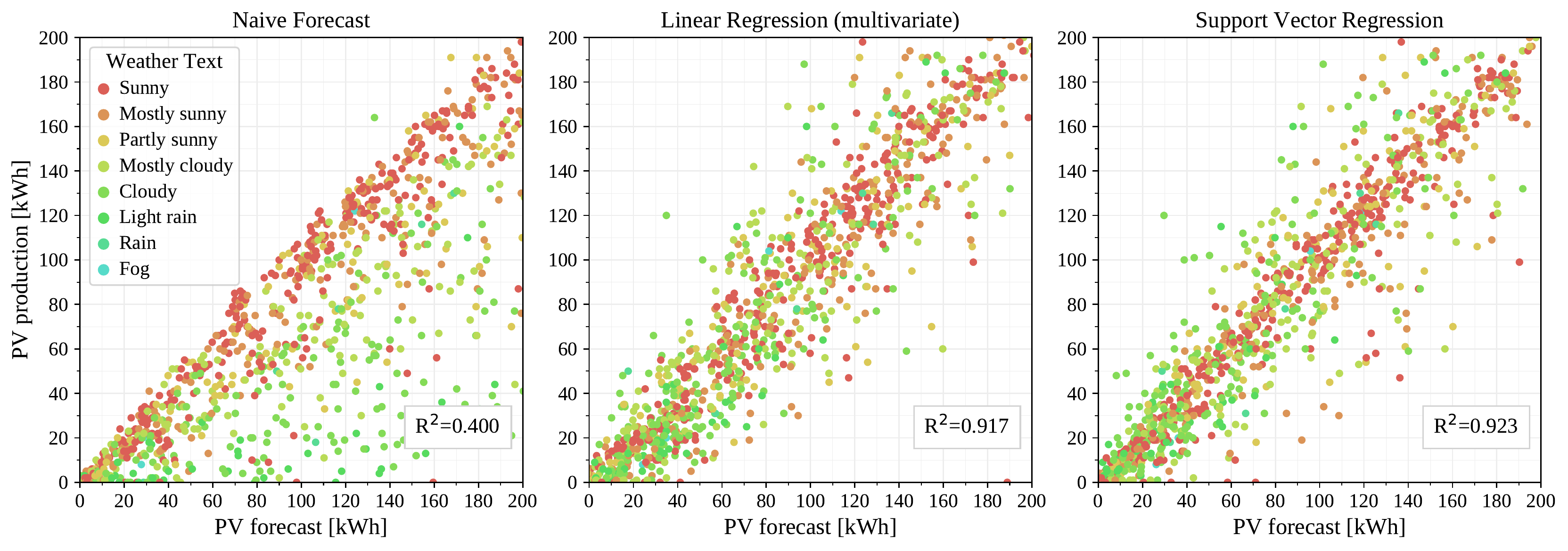}
  \caption{Comparative results of some representative forecasting models: naive model (left), not considering weather conditions; multivariate linear regression (center) and support vector regression (right). 
  The distance to the diagonal is the error between the forecast and the real values. 
  Note that even simple linear models provide accurate forecasts (explained variability R$^2$ above 90\%).}
  \label{fig:f5}
\end{figure*}

\fig{fig:f5}, presents the results of the evaluated "1-hour ahead" regression forecast.
On the left side we can observe the relation between the empirical observation and the predicted value
based only in the solar irradiation, without considering meteorological and other available information (\textit{naive forecast}).
The points that fall below the diagonal are actual values that differ from the expected production due to the impact of short term weather conditions.
The center figure presents the forecast results using a simple multivariate linear regression.
On the right figure, we show the results of the estimation applying the SVR model, fed by the meteorological data,  historical data averages, last production value and clean irradiation
(variables included in \tab{tab:t3}). 
As can be observed, the forecast values are closer to the diagonal. 
Quantitatively, the naive model has a score R$^2$=0.400. In contrast, the SVR achieves R$^2$=0.923, that is, over 90\% of explained variability.
But it is relevant to note that even with a very simple model, without requiring a training phase, it is possible to achieve a good prediction accuracy. 
See for example that the linear model presented achieves an R$^2$=0.917 of explained variability. 

\section{Viability Study}
\label{sec:viability}

In this section we put together the computing and the energy forecast models in order to investigate the viability of the envisaged solution. 
With that aim, we started looking to the power consumption per computing instance $\bar{P}$ in Sect. \ref{sec:computing}, defining $\eta_C$ as the number of instances that can be operated per kW. 
From this parameter, multiplying by a tentative price per instance V$_I$, in \equ{eq:eq1} we derived an expression for the revenue R$_C$ that can be obtained per each kWh. 

\begin{figure}[ht!]
  \centering
  \includegraphics[width=1.0\columnwidth]{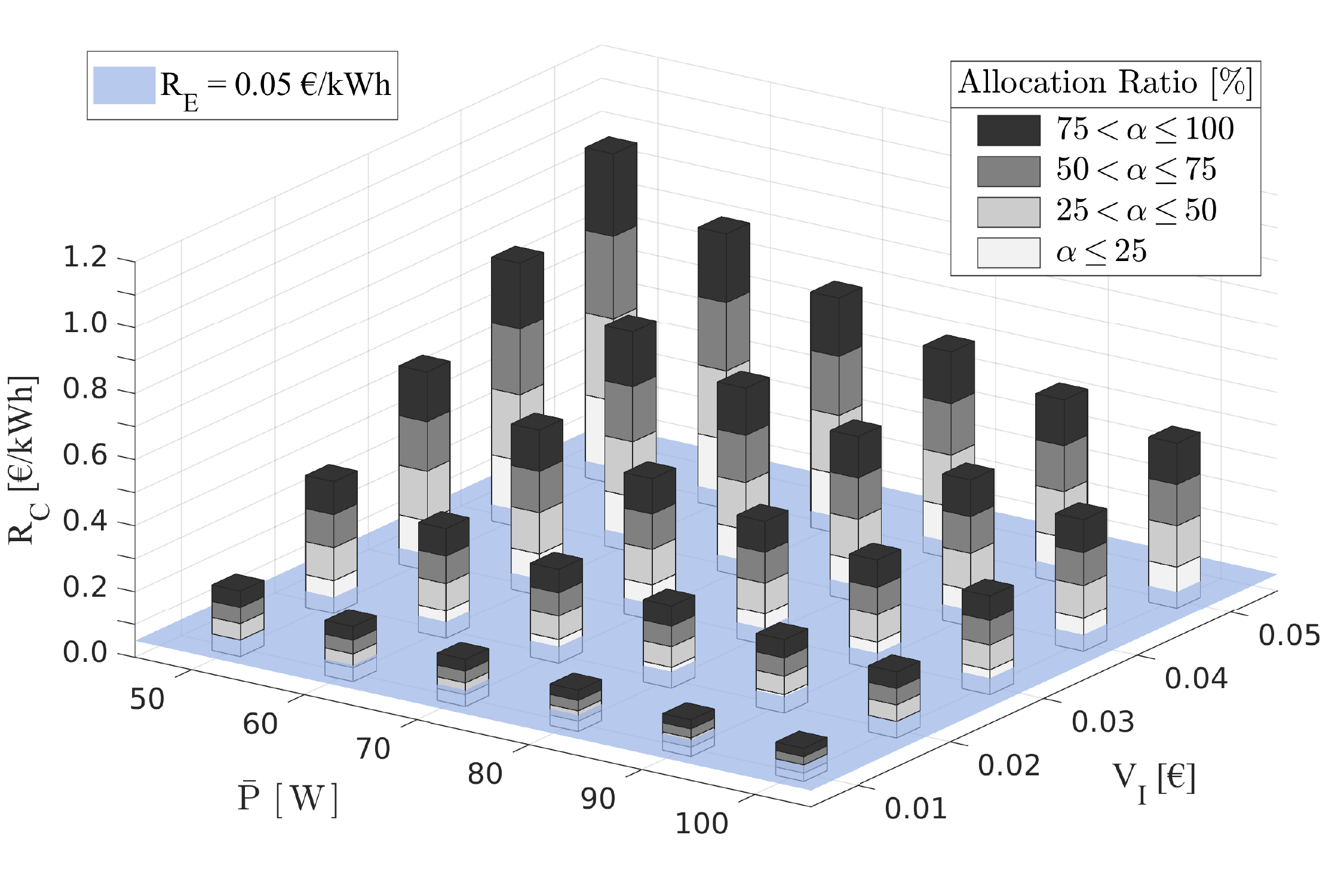}
  \caption{Mapping between power consumption $\bar{\text{P}}$, instance price $V_I$ and potential revenue $R_C$ normalized to 1\,kWh of nominal power installed ($P_{\text{MPP}}$). Bars above the shaded plane represent cases with positive revenues.}
  \label{fig:f6}
\end{figure}

Now, \fig{fig:f6} shows R$_C$ for several values of the variables V$_C$ and $\bar{P}$. 
Consider for example a computing instance (e.g. a t2.xlarge with 4 cores) that features an average power of P=50W ($\eta_C$=20), and match it with a price per instance of V=0.02\,\texteuro/h.
The Z-axis in the plot is indicating the revenue R$_C$ per computed kWh, that is, 
the profit that can be obtained by leasing 20 instances of this type for 1\,hour or, inversely, by leasing one single  instance until 1\,kWh is consumed. 
The scale on the bars represent the allocation ratio $\alpha$, 
plotted in this example for the 75\%, 50\% and 25\% percentiles (recall from Section \ref{sec:computing} that $\alpha$ accounts for the ratio of instances/time actually allocated for computation). 

In the same figure, the flat surface delimits the revenue R$_E$ that could be obtained by injecting to the grid the energy produced (instead of using it for computation) if it was rewarded at 0.05\,\texteuro/kWh by the utility company. 
\equ{eq:eq4} defines the net revenue R$_N$ that can be obtained by using the energy for computation, that is, subtract the revenue R$_E$ from the gross profit R$_C$ generated by allocating computing tasks on computers powered by the same amount of energy. 
This net revenue is interpreted in \fig{fig:f6} as the height of the bars above the grayed surface R$_E$.

\begin{equation}
\label{eq:eq4}
R_N\,[\text{\texteuro}/kWh] = R_C -  R_E
\end{equation}

\fig{fig:f6} makes it clear that even in case of a partial allocation of the computing resources (e.g., due to limited computing demand) 
the option of computing overcomes the revenues obtained by selling the energy to the utility company in most of the cases. 
Following the example of the previous section, when the selling price per instance is 0.02\,\texteuro/h, 
the revenues obtained by leasing a number of instances equivalent to 1\,kWh can overcome the income of selling the energy for any value of the power consumption range, provided that the allocation factor is high enough.
For example, for P$_C$=50W the fraction of actually leased units should be above 12.5\%, while for P$_C$=100W the allocation ratio is required to be above 25\%.

Finally, the payback obtained by adding a computing infrastructure to a PV system can be derived from the net revenue R$_N$ per kWh 
and the total energy $E_T$ produced during the year by the PV system. 
As we have seen, from \equ{eq:eq2} the total energy is parametrized by the equivalent peak-sun-hours PSH per year, 
the nominal PV power installed P$_{\text{MPP}}$ and the PV system losses.
Then, using $E_T$ from \equ{eq:eq2} leads to:
\begin{equation}
\label{eq:eq5}
A\,[\text{\texteuro}/year] = R_N \cdot E_T = R_N \cdot PSH \cdot \eta_{SYS} \cdot P_{MPP}
\end{equation}

Consider the values presented in \tab{tab:t2}, for a particular area (e.g the Mediterranean coast) with 1670.7 PSH, 
take a PV installation with a nominal power of 1\,kWh, and consider that the energy injected is paid back by the utility company at 0.05\,\texteuro/kWh. 
From the example above, 
if we again assume 20 instances consuming 50W each, allocated at a price per instance V$_I$ of 0.02\,\texteuro/h, we obtain a gross revenue R$_C$ = 0.4\,\texteuro~ and net revenue R$_N$ = 0.35\,\texteuro~ per computed kWh. 
From \equ{eq:eq5} and using $\eta_{SYS}$ from \tab{tab:t2} the yearly revenue can be up to 452\,\texteuro~per nominal kW installed when $\alpha=100\%$.   

\begin{figure}[ht!]
  \centering
  \includegraphics[width=1.0\columnwidth]{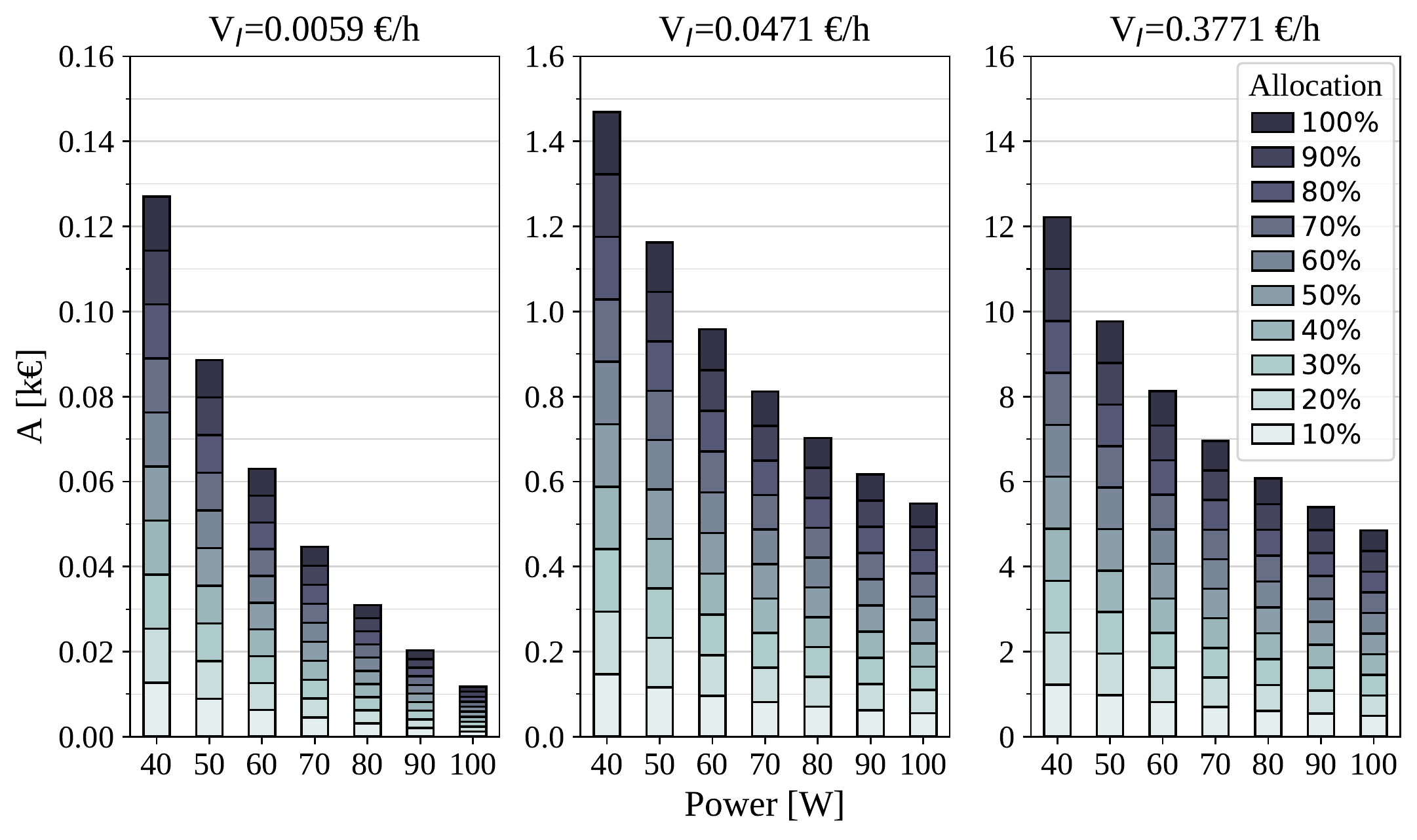}
  \caption{Comparative yearly payback $A$ for different AWS T2 instance prices, according to power consumption and the allocation ratio, i.e percentage of the computational capacity leased. Note the different scales (10x) in the Y-axis.}
  \label{fig:f7}
\end{figure}

In order to get a deeper insight on the revenue range that can be obtained, we evaluate \equ{eq:eq5} considering different values for the power consumption and allocation ratio.
This has been mapped in \fig{fig:f7} to three different prices taken from common AWS instances, which is normalized to 1\,kWh of energy. 
Following the example, taking 20 instances with a power consumption of 50W each, 
computing R$_N$ for a leasing price P=0.0471\,\texteuro/h~per instance (price of the t2.medium) and an allocation ratio of 50\%, 
the annual payback is around 577\,\texteuro~per kWh installed. 
What is important to observe, though, is the huge variation in the revenues due to the instance pricing. 
Given the similar power consumption of very different instances observed in Sec.\ref{sec:computing}, 
the revenue can range from  44\,\texteuro~for a leasing price of 0.0059\,\texteuro/h~(price of the t2.nano) 
to 4843\,\texteuro~for 0.3712\,\texteuro/h~(price of the t2.2xlarge). 
Even considering that the upfront costs of the infrastructure for the two cases would probably be notably different, the revenue disparity is significantly larger. 
This is caused by the 2 orders of magnitude between the prices of the analyzed instances. 
We believe the leasing prices in AWS are not reflecting the value of the hardware.

Another viable alternative is to use virtualization of a single machine (e.g t2.2xlarge with 8 cores) and creating as many instances as cores it has (e.g 8 instances of t2.nano). With this the energy consumption is basically the same as shown in \fig{fig:f2} but the revenue will add up with the number of virtualized instances leased. 

\section{State-of-the-Art, Markets and Applications}
\label{sec:sota}
We have not been able to find in the previous art an answer to the question raised in this study: 
\textit{is it worth launching computation as an alternative to injecting the excess energy to the grid?} 
As we discuss in this section, there are related studies that analyze the operation of computing facilities using renewable energies. 
There is also a significant research effort describing distributed computing facilities for different purposes. 
Nevertheless, as far as we know, the viability of a distributed model on the basis of solar, self-produced energy have not been addressed.


%
The major cloud computing vendors (Amazon, Microsoft and Alphabet Inc.) are firmly supporting sustainable policies to mitigate the energy footprint of their datacenters. 
BloombergNEF recently reported~\cite{BloombergNEF} that the giant cloud-services providers, headed by Google,    
are the corporations closing more power purchase agreements (PPAs) for renewable energy worldwide, demonstrating a clear trend to run cloud datacenters from green energy sources. Sustainability is on the horizon of the key actors. 

Another independent trend on the most representative cloud providers is a certain degree of openness to the edge. 
For example, Microsoft offers the so-called Azure IoT service, on which an agent can be deployed on edge devices (e.g., on-premises servers) enabling them to remotely interact with the Azure cloud hosted services. Amazon IoT Greengrass and Google Cloud IoT Core provide similar functionalities, enabling part of the computation to be moved to the edge. 
For the time being, there is no evidence of synergies between these two trends, namely, 
we are not aware of any efforts from these vendors to sustain edge computation through renewable energy sources.

From another perspective, distributed computing has been deeply studied and adopted in the last 20 years. The concept of distributed clouds is derived from the 2000s peer-to-peer and grid computing communities, 
where computation power was distributed among peers to enable large-scale, massive computations. 
Frameworks such as BOINC \cite{Boinc2004} 
and PlanetLab \cite{PlanetLab2006}, 
among many others, were designed to execute distributed applications using volunteered -altruistic- computing resources,
most of them to support academic, scientific and social goals.

More recently, cryptocurrencies mining (e.g Bitcoin) have exploited this distributed approach to sustain a whole industry, 
relaying on distributed computing peers -miners- connected through an overlay network. 
For example, the "proof of work" concept (first introduced \cite{Dwork1992}) that Bitcoin uses to validate transactions \cite{Gupta2018}, 
consumes vast amounts of computational resources.
Indeed, miners are basically converting energy into cryptocurrency, 
since they are rewarded with a certain amount when successfully adding new blocks to the chain. 
Whether the business is profitable depends on the energy price and the current Bitcoin exchange.

Computing as a service from distributed computation resources have been also exploited in recent initiatives. 
For example, iExec \cite{iExec2020} 
provides a decentralized marketplace for cloud resources, enabling remote computers to join a distributed -although centrally orchestrated- computing pool. 
In this model, cloud resources are traded on a global market, just as another commodity. 
Computing contributors are rewarded when their machines are leased, with a pricing model regulated by offer/demand. 
A similar idea is promoted by P2P-VPS \cite{p2pvps2020}, but in this case rewarded with cryptocurrency
somehow closing the loop.

Interestingly, while we perceive a clear evolution towards decentralization and scaling of computation, 
and we observe notable efforts towards powering data centers using sustainable energy sources, 
we are not observing converging directions to support distributed computation on sustainable small-scale facilities. In addition, the recent emergence of computation marketplaces may be understood as a new trend in the architecture of distributed computing services, which indeed may be boosted by the lower energy costs of self-production. 

\section{Discussion and final remarks}
\label{sec:conclusion}

In this article we have explored the viability to enable computing resources through energy generated by PV producers, providing computation as a service. 
Computing resources can execute edge services analogously to the offered by the cloud service providers (e.g Azure IoT agents). 
Alternatively, those resources can be joined to distributed cloud systems or volunteer computing projects such as those described in the previous section. 

This model offers several advantages to the different players in the ecosystem. 
From a PV producer perspective, it emerges as an alternative to value the exceeding energy, without having to only rely on the utility company to compensate for the injected energy. With this, competition is introduced in the market, favoring the PV producer to shorten the payback of the investment. 

From a cloud computing perspective the model reduces the energetic impact of the technology, relying on small micro-infrastructures spread over the territory and promotes truly green ICT infrastructures.
%
This model may establish synergies with the current cloud computing providers or become an alternative with social, democratic and sustainable values that the involved sectors (energy and cloud computing) cannot guarantee. 
In addition, this model may enable novel computation marketplace initiatives to develop more competitive pricing strategies.

As presented in this article, the idea is technically realizable and viable economic models are possible to motivate its growth, yet technological, social and political uncertainties need to be addressed, including
security and trustworthiness considerations, investment model and regulations, network capacity and data volumes and ownership models for the brokers.

\ifCLASSOPTIONpeerreview

\else


\fi


\IEEEtriggeratref{6}

\bibliography{solar-paper}{}
\bibliographystyle{ieeetr}

\end{document}